\documentclass[aip,reprint,onecolumn]{revtex4-1}

\usepackage{graphicx}
\usepackage{amsmath}
\usepackage{amssymb}
\usepackage{mathrsfs} 					% pour utiliser les lettre "script"

\draft % marks overfull lines with a black rule on the right

\begin{document}

%\preprint{}

\title{Radiative lifetime of the $a\ ^3\Sigma^+$ state of HeH$^+$ from {\it ab~initio} calculations} %Title of paper



%\thanks{}

\author{J. Loreau}
\affiliation{Laboratoire de Chimie Quantique et Photophysique, Universit\'e Libre de Bruxelles, CP160/09 50, av. F. D. Roosevelt, 1050 Bruxelles, Belgium}
\author{J. Li\'evin}
\affiliation{Laboratoire de Chimie Quantique et Photophysique, Universit\'e Libre de Bruxelles, CP160/09 50, av. F. D. Roosevelt, 1050 Bruxelles, Belgium}
\author{N. Vaeck}
\affiliation{Laboratoire de Chimie Quantique et Photophysique, Universit\'e Libre de Bruxelles, CP160/09 50, av. F. D. Roosevelt, 1050 Bruxelles, Belgium}


\date{\today}

\begin{abstract}
The first metastable triplet state of HeH$^+$ was found to be present in ion beam experiments, with its lifetime estimated to be between hundreds of milliseconds and thousand of seconds. In this work, we use {\it ab initio} methods to evaluate the radiative lifetimes of the six vibrational levels of the $a\ ^3\Sigma^+$ of HeH$^+$. The transition $a\ ^3\Sigma^+ \rightarrow X \ ^1\Sigma^+$ is spin-forbidden, but acquires intensity through spin-orbit interaction with the singlet and triplet $\Pi$ states. Large scale CASSCF/MRCI calculations using an adapted basis set were performed to determine the potential energy curves of the relevant states of HeH$^+$ as well as the matrix elements of the dipole and spin-orbit operators. The wave functions and energies of the vibrational levels of the $a\ ^3\Sigma^+$ and $X \ ^1\Sigma^+$ states are obtained using a B-spline method and compared to previous works. We find that radiative lifetime of the vibrational levels increases strongly with $v$, the lifetime of the $v=0$ state being of 150 s. We also analyze the contributions from discrete and continuum parts of the spectrum. With such a long lifetime, the $a\ ^3\Sigma^+$ state could have astrophysical implications.
\end{abstract}


\maketitle %\maketitle must follow title, authors, abstract and \pacs

%\label{}

\section{Introduction}

The helium hydride ion HeH$^+$ is one of the most elementary molecular ions and the first to form in the early universe \cite{Lepp2002}. Due to its high relative abundance, HeH$^+$ was predicted to be observable in astrophysical objects such as planetary \cite{Black1978} and gaseous nebulae, or in metal-poor stars \cite{Harris2004}. However, no infrared emission from HeH$^+$ molecular ion has yet been detected from these objects \cite{Moorehead1988,Liu1997}, although it has been observed in laboratory plasmas for many years \cite{Hogness1925}. 
The formation of HeH$^+$ is mainly due to the radiative association between He and H$^+$ or between He$^+$ and H \cite{Roberge1982}. While it was always supposed that HeH$^+$ formed in its ground $X\ ^1\Sigma^+$ state, one should also consider the possible role of the first metastable triplet state, $a\ ^3\Sigma^+$. This state can indeed be populated and will not decay by collisions if the plasma density is low. As its radiative decay to the ground state is spin-forbidden, it is thus expected to have a very long lifetime.

In studies on the dissociative recombination of HeH$^+$, it was shown that the $a\ ^3\Sigma^+$ state is responsible for a part of the cross section, so that it must be present in the ion beam. The lifetime of this state was postulated by Yousif {\it et al} \cite{Yousif1994} to be longer than the one of He$(1s2s\ ^3S)$, which decays by a relativistic magnetic dipole transition and has a lifetime of approximately 8000 s. In an experiment on the charge-transfer dissociation of HeH$^+$, Strasser {\it et al} \cite{Strasser2000} estimated the lifetime of the triplet state to be in the range of a few hundreds of milliseconds, a much lower value probably due to collisional decay. These two estimations provide lower and upper bounds on the lifetime of this state, but the difference is so large that it motivates a theoretical investigation of the lifetime.

We present in this work the results of large scale {\it ab initio} calculations performed with the MOLPRO program suite \cite{molpro}. These calculations take the spin-orbit coupling between the low-lying singlet and triplet $\Sigma^+$ and $\Pi$ states into account, allowing the spin-forbidden $a\ ^3\Sigma^+ \rightarrow X\ ^1\Sigma^+$ dipole transition to occur. This mechanism was already used to estimate the lifetime of the first triplet state of NO$^+$ \cite{Manaa1991,Palmieri1993}.

The resolution of the vibrational problem in the considered potential energy curves has been done using a B-spline basis set method, which allowed us to estimate the contribution of the continuum states to the lifetime.

\section{Theory}

\subsection{Lifetime}

The inverse of the lifetime $\tau_{i}$ of an initial excited electronic state $\vert i \rangle$ is given in the electric dipole approximation by \cite{Oddershede1979}
\begin{equation}
\tau_{i}^{-1} = \sum_{f} A_{if} = \frac{4 }{3 \hbar c^3} \sum_{f} E_{if}^3 \vert \langle i \vert \boldsymbol \mu  \vert f\rangle \vert ^2
\end{equation}
where the sum extends over all states $\vert f \rangle$ with an energy $E_f < E_i$. The $A_{if}$ are the so-called Einstein coefficients for spontaneous emission and $\boldsymbol \mu$ is the dipole operator. $E_{if}=E_i-E_f$ is the energy difference between states $\vert i \rangle$ and $\vert f \rangle$.

In the Born-Oppenheimer approximation, the total wave function for the initial or the final state is expanded as a product of electronic, vibrational and rotational wave functions. These three types of motions are represented by the quantum numbers $m$ (electronic), $v$ (vibrational) and $J$ (rotational).
In Hund's case (a), the wave function for a state $\vert mvJ \rangle$ is
\begin{equation}
\Psi_{mvJ} = \zeta_{m\Lambda S\Sigma} (R,r) \psi_{mvJ\Lambda S}(R) \mathscr{\bar{D}}^J_{M\Omega}(\theta, \phi)
\end{equation}
$\Lambda$ is a quantum number associated to $L_z$, the projection on the internuclear $z$ axis of the total electronic angular momentum ${\bf L}$. $\Sigma$ is associated to $S_z$, the projection on the $z$ axis of the total electronic spin ${\bf S}$. The total angular momentum is ${\bf J=N+L+S}$, where ${\bf N}$ is the angular momentum for nuclear rotation. $\Omega=\Lambda+\Sigma$ is a quantum number associated with $J_z$.
The vibrational equation that the functions $\psi_{mvJ\Lambda S}(R)$ must satisfy is, in atomic units,
\begin{equation}
\bigg( -\frac{1}{2\mu} \partial^2_R +U_{mvJ\Lambda S}(R) -E_{mvJ\Lambda S} \bigg) \psi_{mvJ\Lambda S}(R) =0
\end{equation}
where $U_{mvJ\Lambda S}(R)$ is the electronic energy, $U_{i\Lambda S}(R)$, corrected by a centrifugal term originating from the rotational Hamiltonian,
\begin{equation}\label{diag_corr}
U_{mvJ\Lambda S}(R) = U_{m\Lambda S}(R) + \frac{1}{2\mu R^2} \Big( J(J+1)-\Omega^2 +S(S+1)-\Sigma^2 \Big)
\end{equation}
and $\mu$ is the reduced mass of the system.
Note that there should be an additional ($L_x^2+L_y^2)$ term due to the diagonal Born-Oppenheimer correction, which we have omitted.

If we neglect the rotational motion, the dipole transition moment between the initial and the final state is then given by 
\begin{equation}
\langle iv \vert \boldsymbol \mu \vert fv^{\prime} \rangle = 
\langle iv\Lambda S \vert \langle i\Lambda S\Sigma\vert \boldsymbol{\mu} \vert f\Lambda^{\prime} S^{\prime}\Sigma^{\prime}\rangle
\vert fv^{\prime}\Lambda^{\prime} S^{\prime} \rangle
\end{equation}
while the lifetime of the vibrational level $v$ of the excited electronic state $i$ is given by
\begin{equation}\label{lifetime}
\tau_{iv}^{-1} = \sum_{f} \sum_{v^{\prime}} A_{iv,fv^{\prime}} \ ,
\end{equation}
where the Einstein coefficients $A_{iv,fv^{\prime}}$ are % idem for J J'
\begin{equation}\label{Einstein_coeff}
A_{iv,fv^{\prime}} = \frac{4 }{3 \hbar c^3} \sum_{f} E_{iv,fv^{\prime}}^3 \vert \langle iv\Lambda S \vert \langle i\Lambda S\Sigma\vert \boldsymbol{\mu} \vert f\Lambda^{\prime} S^{\prime}\Sigma^{\prime}\rangle
\vert fv^{\prime}\Lambda^{\prime} S^{\prime} \rangle
 \vert ^2
\end{equation}

To be exact, the sum over the vibrational levels $v^{\prime}$ of state $f$ in equation (\ref{lifetime}) should be understood as a sum if $v^{\prime}$ is a discrete (bound) level or as an integral if $v^{\prime}$ is a continuum (unbound) level. However, as explained below, we will use a discretization method to treat the continuum, so that the sum will run over the discrete levels until convergence in equation (\ref{lifetime}).
As the ground state is the only state below the $a\ ^3\Sigma^+$ state, the sum over $f$ will reduce to only one term.

To calculate the lifetime of state $\vert i \rangle$, it is thus necessary to know: {\it(i)} its transition moments with all the states $\vert f \rangle$ lower in energy, and {\it(ii)} the vibrational wave functions (and energies) of states $\vert i \rangle$ and $\vert f \rangle$.

\subsection{Spin-orbit coupling}
The transition $a\ ^3\Sigma^+ - X\ ^1\Sigma^+$ is forbidden at all multipole orders due to the spin selection rule $\Delta S =0$. However, it can occur through spin-orbit coupling and the matrix element of interest,
\begin{equation}\label{transition}
\langle X\ ^1\Sigma^+ \vert {\boldsymbol \mu} \vert a\ ^3\Sigma^+ \rangle_{\mathrm{SO}} \ ,
\end{equation}
will be non zero.

We add to the molecular Hamiltonian $H$ the Breit-Pauli $H^{\mathrm{SO}}$ perturbation term, given by \cite{LefebvreBrion2004}
\begin{equation}\label{H_SO}
H^{\mathrm{SO}}=\frac{\alpha^2}{2} \sum_{i,A} \frac{Z_A}{r_{iA}^3}{\boldsymbol l}_{iA}\cdot {\boldsymbol s}_{i}  - \frac{\alpha^2}{2} \sum_{i\neq j} \frac{1}{r_{ij}^3} (\boldsymbol r_{ij}\times \boldsymbol p_{i}) (\boldsymbol s_i + 2 \boldsymbol s_j) \ ,
\end{equation}
where $i$ and $A$ denote electrons and nuclei, respectively and $\alpha$ is the fine structure constant.
The first term in equation (\ref{H_SO}) is the direct spin-orbit interaction, while the second term is the spin-other orbit interaction. As this perturbation mixes the spin and orbital angular momenta of the electrons, the correct quantum number is $\Omega=\Lambda+\Sigma$. In this representation, the states characterized by $\Lambda$ and $\Sigma$ split into $\Omega$ components, whose symmetry can be determined by group theory \cite{Herzberg1966} from the direct product of the spatial and spin symmetry species. The $a\ ^3\Sigma^+$ state will split into two components corresponding to the $\Pi$ and $\Sigma^-$ irreducible representations of the $\mathscr{C}_{\infty v}$ point group. Using standard labeling, these components are denoted $\Omega=1$ and $\Omega=0^-$, respectively. All the $\Omega\neq0$ components are doubly degenerate.
However, the diagonal elements of $H^{\mathrm{SO}}$ can be shown to be proportional to $\Lambda\Sigma$ so that the three components of the $^3\Sigma^+$ state are still degenerate \cite{LefebvreBrion2004}.

The selection rules for spin-orbit coupling are \cite{Kayama1967}
\begin{equation}\label{sel_rules}
\Delta\Omega=0\ ; \quad \Delta S=0,\pm 1\ ; \quad \Delta\Lambda=-\Delta\Sigma=0,\pm1\ ; \qquad \Sigma^+ \longleftrightarrow \Sigma^-
\end{equation}
In accordance with the last rule, there will be no spin-orbit interaction between the $a\ ^3\Sigma^+$ and the $^1\Sigma^+$ states. However, if we take into account higher parts of the spectrum of HeH$^+$, and in particular $\Pi$ states, the transition will become possible. 

If we consider a $^1\Pi$ and a $^3\Pi$ state, we can use the rules (\ref{sel_rules}) to write the $X\ ^1\Sigma^+$ and $a\ ^3\Sigma^+$ wave functions in the spin-orbit representation in terms of the unperturbed functions as
\begin{eqnarray}\label{SO_wf}
&& \vert X \ ^1\Sigma^+_{0^+} \rangle_{\mathrm{SO}}  = c_1 \vert X\ ^1\Sigma^+_{0^+} \rangle + c_2 \vert ^3\Pi_{0^+} \rangle  \nonumber \\ 
&& \vert a\ ^3\Sigma^+_{0^-} \rangle_{\mathrm{SO}}  = c_3 \vert a\ ^3\Sigma^+_{0^-} \rangle + c_4 \vert  ^3\Pi_{0^-} \rangle  \\
&& \vert a\ ^3\Sigma^+_{1} \rangle_{\mathrm{SO}}  = c_5 \vert a\ ^3\Sigma^+_{1} \rangle + c_6 \vert ^1\Pi_{1} \rangle  + c_{7} \vert  ^3\Pi_{1} \rangle  \nonumber
\end{eqnarray} 
where the coefficients $c_i$ are obtained by diagonalizing the spin-orbit matrix in the basis of the unperturbed functions.

The relevant matrix element (\ref{transition}) in the spin-orbit representation can be evaluated, provided that the mixing coefficients and the dipole transition functions are known in the unperturbed basis.
Due to the splitting of the triplet state into two components, the matrix element (\ref{transition}) is split into two parts, according to the value of $\Omega$. However, the matrix element $\langle X\ ^1\Sigma^+_{0^+} \vert \boldsymbol{\mu} \vert a\ ^3\Sigma^+_{0^-} \rangle$ vanishes identically due to the fact that the electric dipole operator cannot connect $0^+$ with $0^-$ states. 
The component $a\ ^3\Sigma^+_{0^-}$ will thus decay through another mechanism such as spin-rotation or relativistic magnetic dipole perturbations, which are smaller by several orders of magnitude.
We will therefore only be able to calculate the lifetime of the $\Omega=1$ component of $a\ ^3\Sigma^+$. However, as the three components are degenerate, the total lifetime of this state will be given by 
\begin{equation}
\tau_{v}^{-1}(a\ ^3\Sigma^+) = \frac{2}{3}\frac{4 }{3 \hbar c^3} \sum_{v^{\prime}} E_{vv^{\prime}}^3 \vert \langle X \ ^1\Sigma^+_{0^+} v \vert \boldsymbol{\mu} \vert a\ ^3\Sigma^+_1 v^{\prime} \rangle_{\mathrm{SO}} \vert ^2
\end{equation}

The matrix element (\ref{transition}) can be evaluated for the $\Omega=1$ component using (\ref{SO_wf}) as
\begin{equation}\label{SO_mat_el}
\langle X \ ^1\Sigma^+_{0^+} \vert \boldsymbol{\mu} \vert a\ ^3\Sigma^+_{1} \rangle_{\mathrm{SO}} = c_1c_6 \langle X \ ^1\Sigma^+ \vert \boldsymbol{\mu} \vert ^1\Pi\rangle + c_2c_5 \langle  ^3\Pi \vert \boldsymbol{\mu} \vert a\ ^3\Sigma^+ \rangle + c_2c_{7} \langle  ^3\Pi \vert \boldsymbol{\mu} \vert ^3\Pi \rangle  
\end{equation}

All the dipole matrix element occurring in equation (\ref{SO_mat_el}) are non zero for the $\mu_x$ and $\mu_y$ components of the dipole operator, and the extension of equations (\ref{SO_wf}) and (\ref{SO_mat_el}) to the case of more than one singlet or/and triplet $\Pi$ state is straightforward.

\section{Results and discussion}

\subsection{Dipole transition function}

\begin{figure}[h!]
\centering
\hspace{-1cm}
\includegraphics[angle=-90,width=13cm]{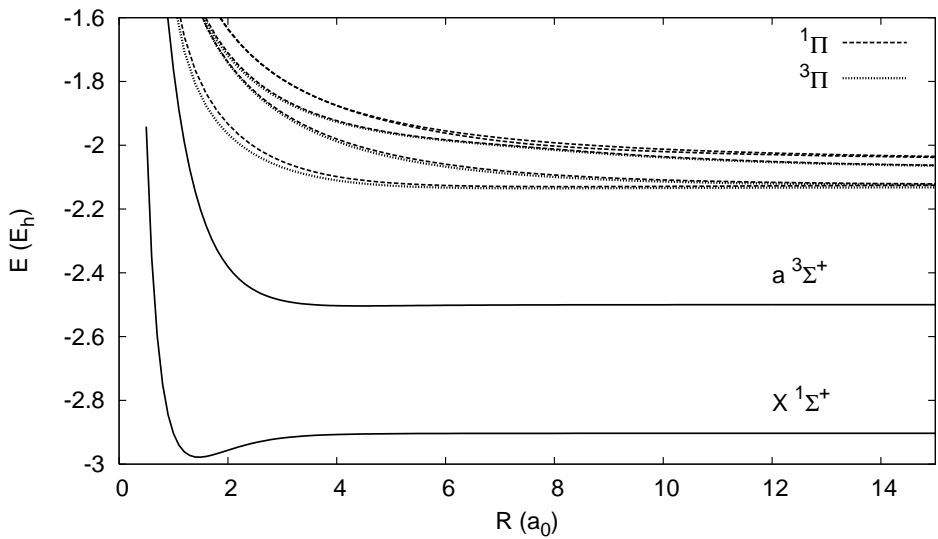}
\caption{\label{states_fig}Potential energy curves of the molecular states included in the calculations.}
\end{figure}

We will consider here the lower part of the spectrum of HeH$^+$, which is composed of states that dissociates either into H$^+$ + He($1snl\ ^{1,3}L$) or into H($nl$) + He$^+(1s)$. 
As the various $^1\Sigma^+$ and $^3\Sigma^+$ states cannot interact through spin-orbit perturbation, it is not necessary to include in the spin-orbit calculations other $\Sigma$ states than the $X\ ^1\Sigma^+$ and the $a\ ^3\Sigma^+$.
On the other hand, all the singlet and triplet $\Pi$ states will contribute to the matrix element (\ref{transition}). We have found (see below) that five $^1\Pi$ and three $^3\Pi$ are sufficient to describe correctly the transition. The states included are given in table \ref{states_tab} together with their dissociation products, and their adiabatic potential energy curves are presented in figure \ref{states_fig}.

All the calculations were done using the MOLPRO program \cite{molpro} and an adapted basis set which consists for each atom of the aug-cc-pV5Z (or AV5Z) basis set \cite{Dunning1989,Woon1994} augmented by [$3s$, $3p$, $2d$, $1f$] Gaussian-type orbitals optimized to reproduce the spectroscopic orbitals of the He and H excited states (see \cite{Loreau2010} for details).
To obtain the potential energy curves for the electronic states, we performed a state-averaged CASSCF \cite{Werner1985,Knowles1985} followed by a CI calculation using an active space of five $\sigma$, ten $\pi$ and one $\delta$ orbitals. The Breit-Pauli spin-orbit matrix has been calculated on the basis of the unperturbed CASSCF eigenfunctions. To include additional correlation effects, we replaced the CASSCF energies (the diagonal elements of the spin-orbit matrix) by the calculated CI energies. This matrix is then diagonalized using the state interacting method \cite{molpro_SO} implemented in MOLPRO. The diagonalization of the spin-orbit matrix corresponding to all ($\Lambda,\Sigma$) states of table \ref{states_tab} provides 32 roots corresponding to 4, 4, 9 and 3 states with $\Omega=0^+, 0^-,1$ and 2, respectively. 

\begin{table}[htdp]
\begin{center}\begin{tabular}{cccc}
 Symmetry & $U_{i\Lambda S}(R=70)$ & Dissociative atomic states\vspace{1mm} \\ \hline 
 $X\ ^1\Sigma^+$	& -2.90324307 & H$^+$ + He($1s^2\ ^1S$) \\ [0.3ex]
 $a\ ^3\Sigma^+$	& -2.49996040 & H($1s$) + He$^+(1s)$ \\ [0.3ex]
 $1\ ^1\Pi$		& -2.12491660 & H($2p$) + He$^+(1s)$ \\ [0.3ex]
 $2\ ^1\Pi$		& -2.12368473 & H$^+$ + He($1s2p\ ^1P^{o}$)  \\  [0.3ex]
 $3\ ^1\Pi$		& -2.05639837 & $\frac{1}{\sqrt{2}}$H($3p$) + $\frac{1}{\sqrt{2}}$H($3d$) + He$^+(1s)$ \\ [0.3ex]
 $4\ ^1\Pi$		& -2.05616425 & H$^+$ + He($1s3d\ ^1D$) \\ [0.3ex]
 $5\ ^1\Pi$		& -2.05456333 & $\frac{1}{\sqrt{2}}$H($3p$) - $\frac{1}{\sqrt{2}}$H($3d$) + He$^+(1s)$ \\ [0.3ex]
 $1\ ^3\Pi$		& -2.13282525 & H$^+$ + He($1s2p\ ^3P^{o}$) \\ [0.3ex]
 $2\ ^3\Pi$		& -2.12491845 & H($2p$) + He$^+(1s)$ \\  [0.3ex]
 $3\ ^3\Pi$		& -2.05814410 & H$^+$ + He($1s3p\ ^3P^{o}$) \\ [0.3ex]
\end{tabular} \caption{\label{states_tab} Molecular states included in the calculations, together with their dissociative products and energy at $R=70$ au.}
\end{center}
\end{table}

\begin{figure}[h!]
\centering
\hspace{-1cm}
\includegraphics[angle=-90,width=13cm]{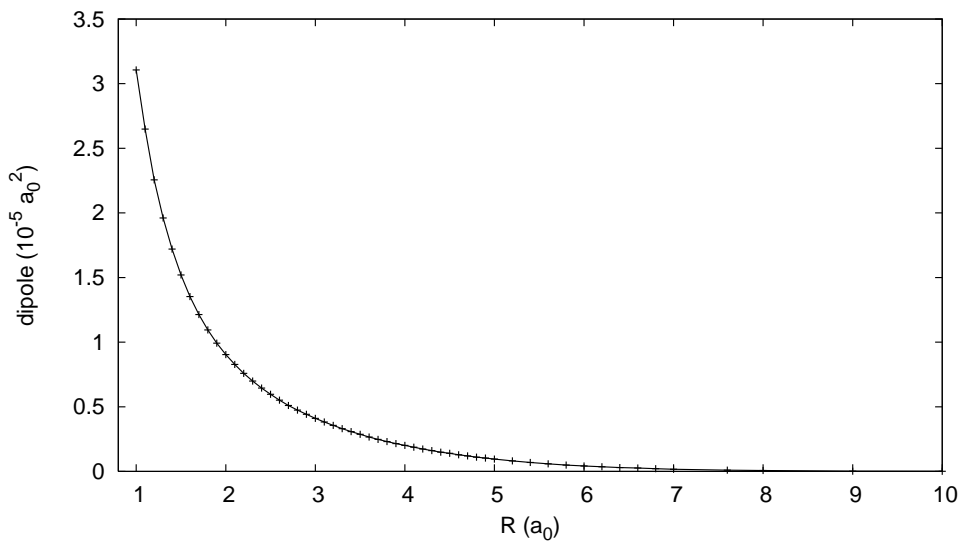}
\caption{\label{dipSO}Dipole transition function $\langle X \ ^1\Sigma^+_{0^+} \vert \boldsymbol{\mu} \vert a\ ^3\Sigma^+_1 \rangle_{\mathrm{SO}}$.}
\end{figure}

The evolution of the dipole transition moment (\ref{transition}) as a function of the internuclear distance is presented in figure \ref{dipSO}. It is of order of $10^{-5}$ $a_0^2$, in agreement with the fact that the spin-orbit interactions are small in light molecules. It also decreases rapidly, corresponding to the fact that the value of the spin-orbit matrix elements under consideration vanish in the atomic limit.

\begin{figure}[h!]
\centering
\hspace{-1cm}
\includegraphics[angle=-90,width=13cm]{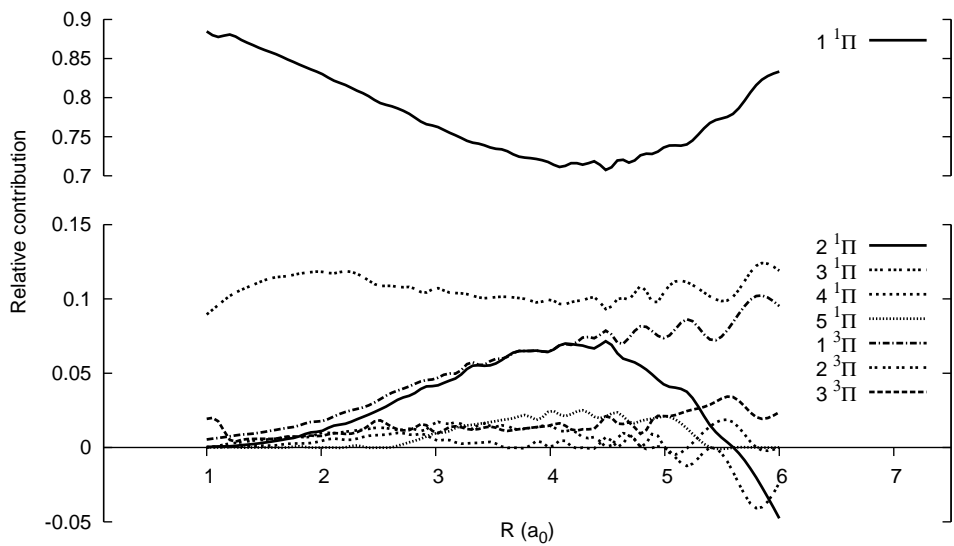}
\caption{\label{dipSO_rel}Relative contributions of the five $^1\Pi$ and the three $^3\Pi$ states to the dipole transition function $\langle X \ ^1\Sigma^+_{0^+} \vert \boldsymbol{\mu} \vert a\ ^3\Sigma^+_1 \rangle_{\mathrm{SO}}$.}
\end{figure}

On figure \ref{dipSO_rel} are presented the weight of each of the $\Pi$ states taken into account into the dipole matrix element $\langle X \ ^1\Sigma^+_{0^+} v \vert \boldsymbol{\mu} \vert a\ ^3\Sigma^+_1 v^{\prime} \rangle_{\mathrm{SO}}$. We see that the matrix element is dominated by the contribution from the first $^1\Pi$ state. The inclusion of the second and third $^1\Pi$ states modify the matrix element by more than 10 \%, but the fourth and fifth $^1\Pi$ states bring an additional correction of only 2--3\% so that it can be supposed that the inclusion of more singlet $\Pi$ states will not affect dramatically the lifetime. The contribution of the first triplet $\Pi$ state increases with the internuclear distance but does not exceed 10\%, while the second and third $^3\Pi$ states contribute to less than 3\%. It is therefore reasonable to assume that the value of the lifetime obtained when considering five $^1\Pi$ and three $^3\Pi$ states is correct up to a few percents.

\subsection{Vibrational analysis of the $X\ ^1\Sigma^+$ and $a\ ^3\Sigma^+$ states}

The vibrational analysis of the $X\ ^1\Sigma^+$ and $a\ ^3\Sigma^+$ states was done using a B-spline basis set method \cite{deBoor1978}.

The {\it ab initio} calculation of molecular vibrational spectra involving highly excited vibrational states implies the variational resolution of the purely vibrational Schr\"odinger equation by analytical FBS (Finite Basis Set) or numerical DVR (Discrete Variable Representation) approaches. %The size of the matrices to be diagonalized in both cases rapidly blows up when the molecular size grows. 
In this work, B-spline basis sets are used as alternative FBS method. The flexibility of B-splines has been demonstrated in atomic and molecular calculations \cite{Shore1973, Bachau2001} by an accurate description of both the bound and the continuum states and their efficiency in the resolution of the nuclear motion of molecules has been assessed. In opposition with the DVR, which is constrained by a uniform distribution of the grid points, the B-splines allow more flexibility from the definition of different cavities in which the number of grid points can be adjusted.

The definition of a B-spline basis set on a cavity of size $L=x_{\textrm{min}}-x_{\textrm{max}}$ starts with the definition of a sequence of $N$ real-valued knots $\{{t_i}\}$, satisfying $t_i\leq t_{i+1}$. Using this knot sequence, a set of $N$ polynomials, all with same degree, is defined: the B-splines of order $K$ (and degree $K-1$). The recursive definition of the B-splines $B_{i,k}(x)$ is the following:
\begin{eqnarray}
 B_{i,1}(x) & =  & \left\{ \begin{array}{ll}
                          1  \qquad \textrm{if} \quad t_i \leq x \leq t_{i+1} \\   
                          0  \qquad \textrm{otherwise}
                        \end{array} \right.         \\ 
B_{i,k}(x) & = & \frac{x-t_i}{t_{i+k-1}-t_i} B_{i,k-1}(x) + \frac{t_{i+k}-x}{t_{i+k}-t_{i+1}} B_{i+1,k-1}(x)
\end{eqnarray}

It should be mentioned that we are free to choose the order $K$, the knot sequence and the interval of the basis set. The space that is spanned by a B-spline depends on this degree; the higher the degree, the larger the space that is spanned.
We used B-splines of order 13 on a grid with $x_{\textrm{min}}=0.5$ au and $x_{\textrm{max}}=100$ au divided by 800 equidistant knot points.

Extensive studies of the ground state of HeH$^+$ have been done by Kolos and Peek \cite{Kolos1976a} or Bishop and Cheung \cite{Bishop1979}. We reproduce the value for the position of the minimum found by these authors, which is 1.463 au. Kolos and Peek find a dissociation energy of 16455.64 cm$^{-1}$ and Bishop and Cheung obtained a value of 16456.15 cm$^{-1}$ when Born-Oppenheimer diagonal corrections  are included. Our result of 16465.5 cm$^{-1}$ is therefore about 9 cm$^{-1}$  larger. The replacement of the AV5Z by the AV6Z basis set leads to an improvement of less than 2 cm$^{-1}$.

The vibrational energies of the ground state of HeH$^+$ have been studied before using diagonal Born-Oppenheimer corrections by Bishop and Cheung \cite{Bishop1979}, or with non-adiabatic and relativistic effects by Stanke {\it et al} \cite{Stanke2006}. As was found in these papers, we obtain 12 vibrational levels. Although we do not reproduce the absolute energies of the levels, the energy separation between these levels is correct with an error of less than 3 cm$^{-1}$, as shown in table \ref{vib_X_3}. The $v=11$ level, not shown in table \ref{vib_X_3}, has a binding energy of 1.35 cm$^{-1}$, to be compared with he value of 1.30 cm$^{-1}$ found by Stanke {\it et al}. It should be noted that the use of the AV6Z instead of the AV5Z basis set leads to different absolute energies for the vibrational levels, but that the spacing remains the same.

%\subsection{$a\ ^3\Sigma^+$ state}
The $a\ ^3\Sigma^+$ state has been studied by Michels \cite{Michels1966}, and a more accurate study was done by Kolos \cite{Kolos1976b} using variational wave functions in elliptic coordinates. Both authors found the minimum to be located at 4.47 au, but the dissociation energy found by Michels is 661.4 cm$^{-1}$ while the result of Kolos is 849.0 cm$^{-1}$. We find an equilibrium position of 4.452 au, and a dissociation energy of 849.79 cm$^{-1}$.

The only calculations on the vibrational levels of the $a\ ^3\Sigma^+$ state were done using the potential energy curve given by Michels completed by an analytical expression at large $R$ \cite{Yousif1994} and it was found that this state supports five vibrational levels. The binding energies of the levels $v=0-3$ can be found in this article and in \cite{Chibisov1996} where more precise values are presented. We reproduce the energy of the levels up to 1 cm$^{-1}$.
However, using our potential energy curve we find that this state supports six vibrational levels, with the $v=5$ state being bound by less than 2 cm$^{-1}$. The total energies are given in table \ref{vib_a}, together with the binding energies.

The vibrational energies presented in table \ref{vib_a} correspond to Hund's case (b) for $J=0$ so that the centrifugal correction in equation (\ref{diag_corr}) vanishes.
As we have seen, we will only be able to calculate the lifetime of the $\Omega=1$ components of the $a\ ^3\Sigma^+$ state. For these, the centrifugal term also vanishes (see equation (\ref{diag_corr})), so that the vibrational energies presented in table \ref {vib_a} energies are still valid.

\begin{table}[htdp]
\begin{center}\begin{tabular}{crrrr}
$v$ & Basis 1 & Basis 2 & Stanke {\it et al} \cite{Stanke2006} & Bishop and Cheung \cite{Bishop1979} \\ [0.3ex] \hline 
0	& 2910.40		& 2910.57		& 2911.02		& 2911.29 \\
1	& 2604.12		& 2604.13		& 2604.21		& 2604.32 \\
2	& 2296.13		& 2296.02		& 2295.64		&  \\
3	& 1983.20		& 1983.02		& 1982.13		&  \\
4	& 1662.19		& 1661.90		& 1660.45		&  \\
5	& 1330.39		& 1329.98		& 1327.91		&  \\
6	& 987.71		& 987.20		& 984.50		&  \\
7	& 643.06		& 642.45		& 639.35		&  \\
8	& 330.72		& 330.26		& 327.49		&  \\
9	& 117.51		& 117.61		& 116.22		&  \\
10	& 24.92		& 24.88		& 24.44		&  \\
\end{tabular} \caption{\label{vib_X_3}Energy difference $E_v-E_{v+1}$ between two successive vibrational levels of the $X\ ^1\Sigma^+$ state of HeH$^+$ in cm$^{-1}$ and comparison with previous works. Basis 1: AV5Z + adapted basis set; Basis 2: AV6Z + adapted basis set.}
\end{center}
\end{table}

% provient de tripsig1_3 R=1-100
\begin{table}[htdp]
\begin{center}\begin{tabular}{ccrr}
$v$ & $E_{\mathrm{tot}}$ & $E_{\mathrm{bind}}$ & $E_v-E_{v+1}$ \\ [0.3ex] \hline 
0	& -2.502 969 82	& -664.834	& 297.67	 \\
1	& -2.501 613 56	& -367.169	& 199.92	 \\
2	& -2.500 702 66	& -167.249	& 109.99	 \\
3	& -2.500 201 52	& -57.260		& 44.61	 \\
4	& -2.499 998 26	& -12.652		& 11.53	 \\
5	& -2.499 945 75	& -1.126		& 	 \\
Unbound & -2.499 940 62	& 
\end{tabular} \caption{\label{vib_a}Total and binding energies of the bound vibrational levels of the $a\ ^3\Sigma^+$ state of HeH$^+$, as well as the energy difference $E_v-E_{v+1}$ between two successive vibrational levels. Total energies in atomic units, binding energies in cm$^{-1}$. }
\end{center}
\end{table}

\begin{figure}[h!]
\centering
\hspace{-1cm}
\includegraphics[angle=-90,width=10cm]{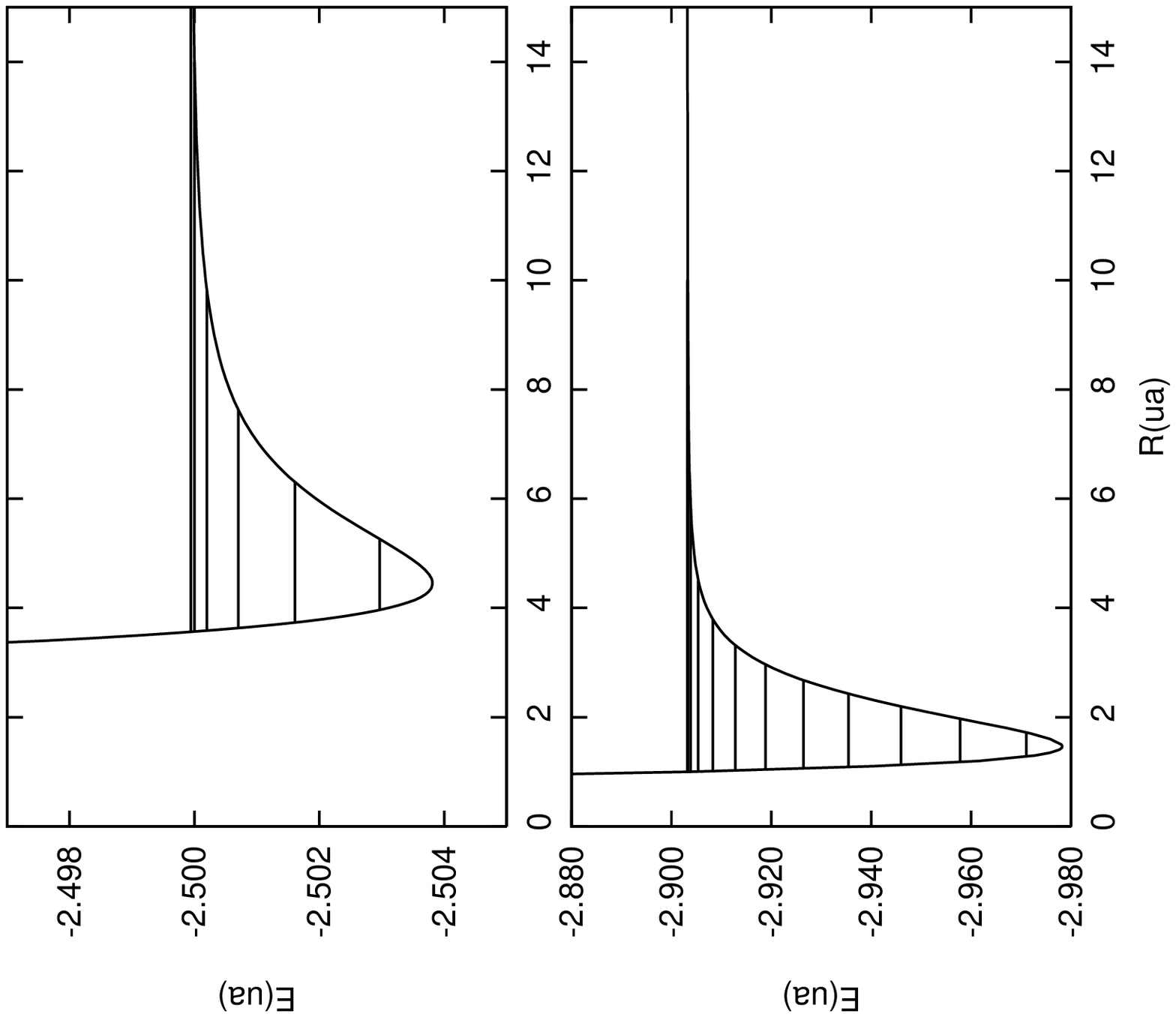}
\caption{\label{compXa} Adiabatic potential energy curves of the first triplet $a\ ^3\Sigma^+$ state (above) and of the ground $X\ ^1\Sigma^+$ state (below) of HeH$^+$ and position of the bound vibrational levels supported by these states.}
\end{figure}

\subsection{Calculation of the lifetime}

As can be seen from equation (\ref{Einstein_coeff}), the Einstein coefficients will depend on the dipole matrix element between the initial and final electronic state, and on the overlap of the vibrational functions. We have seen that the dipole is very small, but in addition the overlap of the vibrational functions is poor since the two states under consideration have their minima located at very different geometries, as illustrated on figure \ref{compXa}.

Using the B-splines, it is a simple task to integrate the overlap between the vibrational functions, multiplied by the dipole matrix element. We performed the integration on a grid $R\in [1,100]$. For the continuum part of the spectrum, it is necessary to sum over all vibrational functions of the pseudo continuum spectrum until convergence.

\begin{figure}[h!]
\centering
\hspace{-1cm}
\includegraphics[angle=-90,width=11cm]{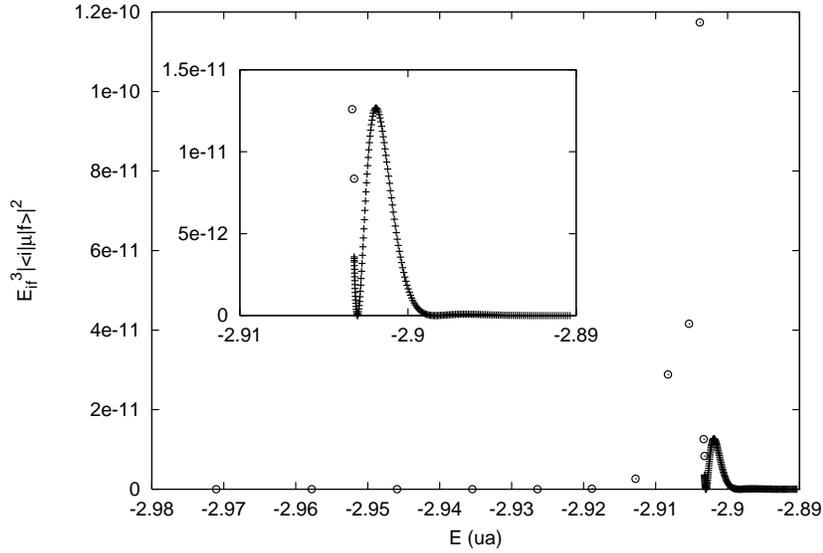}
\caption{\label{A_if}Contribution from the discrete (circles) and continuum (crosses) parts of the vibrational spectrum to the Einstein coefficients $A_{if}$ for $v_i=1$ as a function of the energy of the final vibrational levels.}
\end{figure}

On figure \ref{A_if} are presented the contributions from the discrete and continuum parts of the vibrational spectrum to the Einstein coefficients $A_{if}$. It is seen that the convergence is reached with 200 continuum functions. To represent this contribution, it is necessary to take into account the fact that the vibrational continuum has been discretized by multiplying the $A_{if}$ by the density of states $\rho(E)=2/(E(v_{f+1})-E(v_{f-1}))$. For the discrete part of the spectrum, there is no density of states, but the $A_{if}$ should still be multiplied by some value to allow comparison since $\rho(E)$ depends on the energy units we choose. Following \cite{Chibisov1996}, we use a density for the bound states given by $\rho(E)=1/(E(v_{f+1})-E(v_{f}))$. On figure \ref{A_if}, we observe the continuity of the results around the dissociation limit.

The lifetime, as well as the relative contribution from bound and continuum states, are presented in table \ref{lifetime_result}. We observe that the lifetime increases with the vibrational number $v$. We also see that the contribution of the continuum to the lifetime is very small (5\%) for $v_i=0$, but increases with the value of $v_i$ up to 30\% for $v_i=5$. The contribution of the higher lying $\Pi$ states will probably reduce the lifetime by a few percent, but our calculations provide an upper bound of 150 s on the lifetime of the $v=0$ level of the $a\ ^3\Sigma^+$ state in the absence of collisional decay.

 \begin{table}[htdp]
\begin{center}\begin{tabular}{cccc}
 $v_{i}$	& Bound & Continuum & Lifetime (seconds) \\[0.5ex] \hline 
0	& 0.949	& 0.051	& 149 		\hspace{1cm} \\
1	& 0.906	& 0.094	& 211 		\hspace{1cm} \\
2	& 0.786	& 0.214	& 347 	\hspace{1cm} \\
3	& 0.709	& 0.291	& 738 	\hspace{1cm} \\
4	& 0.694	& 0.306	& 2288 	\hspace{1cm} \\
5	& 0.692	& 0.308	& 14267 	\hspace{1cm} \\
\end{tabular} \caption{\label{lifetime_result}Relative contributions of the discrete and of the continuum part of the vibrational spectrum to the sum of the Einstein coefficients $A_{if}$ and value of the lifetime of the $\Omega=\pm1$ component of $a\ ^3\Sigma^+$ for $v_i=0-5$.}
\end{center}
\end{table}

\section{Conclusions}

We have determined the radiative lifetime of the $a\ ^3\Sigma^+$ state of HeH$^+$ using {\it ab initio} methods. The decay of this state onto the ground $X\ ^1\Sigma^+$ state is spin-forbidden but can occur through spin-orbit coupling. We took into account the interaction of the $a\ ^3\Sigma^+$ and $X\ ^1\Sigma^+$ states with the first five $^1\Pi$ and three $^3\Pi$ states of HeH$^+$ to estimate the dipole transition matrix element $\langle X\ ^1\Sigma^+ \vert {\boldsymbol \mu} \vert a\ ^3\Sigma^+ \rangle_{\mathrm{SO}}$. The vibrational energies and wave functions of the $a\ ^3\Sigma^+$ and $X\ ^1\Sigma^+$ states were obtained using a B-spline method and were found to agree well with previous calculations. We presented theoretical values of the lifetime of the six vibrational levels of the $a\ ^3\Sigma^+$ state. The lifetime is found to be of about 150 s for the $v=0$ state and increases rapidly with $v$, as does the contribution of the continuum states to the lifetime. 
Such a long lifetime suggests that HeH$^+$ could be present in the $a\ ^3\Sigma^+$ state in astrophysical environments. We will investigate the radiative association in this state in a separate work.

\begin{acknowledgments}
J. Loreau would like to thank the FRIA for financial support. This work was supported by the Fonds National de la Recherche Scientifique (IISN projects) and by the ÒAction de Recherche Concert\'eeÓ ATMOS de la Communaut\'e Fran\c caise de Belgique.
\end{acknowledgments}

% If in two-column mode, this environment will change to single-column format so that long equations can be displayed. 
%\begin{widetext}
%$$\mbox{put long equation here}$$
%\end{widetext}

% Figures should be put into the text as floats. 
% Use the graphics or graphicx packages (distributed with LaTeX2e).
% Put the label that you will use with \ref{} command in the braces of the \label{} command.
%
% \begin{figure}
% \end{figure}

% \begin{table}
% \end{tabular}
% \end{table}



\end{document}